\documentclass[twocolumn]{aastex63}
\usepackage{tabularx}
\usepackage{chngcntr}
\usepackage{mathtools,amssymb}
\usepackage{lineno}
\usepackage{comment}

\shorttitle{Bridging the gap}
\shortauthors{Lu et al.}
\graphicspath{{./}{figures/}}

\begin{document}

\title{Bridging the gap --- the disappearance of the intermediate period gap for fully convective stars, uncovered by new ZTF rotation periods}

\correspondingauthor{Yuxi(Lucy) Lu}
\email{lucylulu12311@gmail.com}

\newcommand{\ith}{\ensuremath{^{\rm th}}}
\newcommand{\teff}{\ensuremath{T_{\mbox{\scriptsize eff}}}}
\newcommand{\kepler}{{Kepler}}
\newcommand{\gaia}{{Gaia}}
\newcommand{\tess}{{TESS}}

\newcommand{\prot}{\ensuremath{P_{\mbox{\scriptsize rot}}}}
\newcommand{\pgbr}{\ensuremath{(G_{\rm BP} - G_{\rm RP})}} 
\newcommand{\bprp}{\ensuremath{G_{\rm BP} - G_{\rm RP}}}
\newcommand{\rper}{\ensuremath{R_{\mbox{\scriptsize per}}}}
\newcommand{\ro}{\ensuremath{Ro}}
\newcommand{\msun}{\ensuremath{\mbox{M}_{\odot}}}

\newcommand{\racomment}[1]{{\color{blue}#1}}

\newcommand{\amnh}{American Museum of Natural History, Central Park West, Manhattan, NY, USA}
\newcommand{\cca}{Center for Computational Astrophysics, Flatiron Institute, 162 5\ith\ Avenue, Manhattan, NY, USA}
\newcommand{\columbia}{Department of Astronomy, Columbia University, 550 West 120\ith\ Street, New York, NY, USA}

\author[0000-0003-4769-3273]{Yuxi(Lucy) Lu}
\affiliation{\columbia}
\affiliation{\amnh}

\author[0000-0002-2792-134X]{Jason L.\ Curtis}
\affiliation{\columbia}

\author[0000-0003-4540-5661]{Ruth Angus}
\affiliation{\amnh}
\affiliation{\cca}
\affiliation{\columbia}

\author[0000-0001-6534-6246]{Trevor J.\ David}
\affiliation{\cca}
\affiliation{\amnh}

\author[0000-0002-0842-863X]{Soichiro Hattori}
\affiliation{\columbia}



\begin{abstract}
The intermediate period gap, discovered by Kepler, is an observed dearth of stellar rotation periods in the temperature--period diagram at $\sim$ 20 days for G dwarfs and up to $\sim$ 30 days for early-M dwarfs.
However, because Kepler mainly targeted solar-like stars, there is a lack of measured periods for M dwarfs, especially those at the fully convective limit.
Therefore it is unclear if the intermediate period gap exists for mid- to late-M dwarfs.
Here, we present a period catalog containing 40,553 rotation periods (9,535 periods $>$ 10 days), measured using the Zwicky Transient Facility (ZTF).
To measure these periods, we developed a simple pipeline that improves directly on the ZTF archival light curves and reduces the photometric scatter by 26\%, on average.
This new catalog spans a range of stellar temperatures that connect samples from Kepler with MEarth, a ground-based time domain survey of bright M-dwarfs, and reveals that the intermediate period gap closes at the theoretically predicted location of the fully convective boundary ($G_{\rm BP} - G_{\rm RP} \sim 2.45$ mag).
This result supports the hypothesis that the gap is caused by core--envelope interactions.
Using gyro-kinematic ages, we also find a potential rapid spin-down of stars across this period gap.
\end{abstract}

\keywords{Stellar rotation (1629) --- Catalogs (205) --- Lomb--Scargle periodogram (1959) --- Period search (1955)}


\section{Introduction}\label{sec:intro}
Sun-like stars spin down as approximately $t^{1/2}$ during their time on the main sequence.
This relation, referred to as ``Skumanich-style spin-down,'' was first pointed out observationally by \cite{Skumanich1972} based on the Sun and a few young star clusters.
This law was then derived later in theoretical studies based on various observations \citep[e.g.][]{Kraft1967, Kawaler1988, Krishnamurthi1997}. 
In the early 2000s, inspired by this law,  \cite{Barnes2003} established \textit{gyrochronology}, the method of age-dating stars based on their rotation periods and colors. 
Other age-dating techniques, such as isochrone age-dating or asteroseismology, are not currently able to provide accurate or precise ages
for low-mass main-sequence stars.
The degeneracy of isochrones in the Hertzsprung--Russell diagram can induce age uncertainties of more than 100\% for late-K to M dwarfs, and no asteroseismic signals have been detected in these stars \citep[for an overview of stellar ages, we refer the readers to][]{Soderblom2010}. 
As more age-dating methods are tested to their limits, gyrochronology seems to be one of the best methods to obtain precise and accurate ages for individual low-mass, main-sequence field stars.
Even better, observations suggest that K and M dwarfs lose angular momentum through magnetic wind braking faster than F and G stars, making them the most suitable age-dating targets with gyrochronology.

Unfortunately, due to the chaotic nature of magnetic fields in stars, many deviations from the $t^{1/2}$ Skumanich law that are not well understood have been discovered in observations.
Therefore, detailed stellar spin-down models \citep[e.g.][]{Barnes2010, MacGregor1991, Brown2014, Matt2015, vansaders2016, Spada2020} mostly rely on empirical calibrations, which require large numbers of period measurements.
Fortunately, with recent large space-based photometric surveys such as \kepler\ \citep{Borucki2010}, K2 \citep{Howell2014}, and TESS \citep{TESS}, numerous rotation period (\prot) catalogs \citep[e.g.][Holcomb et al. in prep.]{McQuillan2013, McQuillan2014, Garcia2014, Santos2019, Santos2021, Gordon2021} with precise position and kinematic information provided by \gaia\ \citep{gaia2018} are now available, and we are now able to explore the field of gyrochronology in greater detail.
With these catalogs, many methods, such as asteroseismology \citep[e.g.][]{Angus2015, Hall2021}, open clusters \citep[e.g.][]{Curtis2020, Agueros2018}, binary stars (Godoy-Rivera et al. in prep.), and kinematic ages \cite[][]{Angus2020, Lu2020}, have been used to calibrate gyrochronology relations.
However, since Kepler mainly targeted solar-like stars, we still lack period calibrators at the low-mass end, where gyrochronology is perhaps most applicable. 

Pushing the rotation period sample for stars towards lower masses and encompassing the fully convective limit is not only important for calibrating gyrochronology, but also crucial for understanding the physical mechanisms responsible for K and M dwarf spin-down. 
One possible deviation from the $t^{1/2}$ law is the peculiar period dearth in temperature--rotation period space for low-mass stars first discovered with Kepler by \cite{McQuillan2013}. 
This dearth, sometimes called the intermediate period gap, represents a valley of period detection in temperature--period space, starting from $\sim$20 days for G dwarfs and up to $\sim$30 days for early-M dwarfs (shown later in Figure~\ref{fig:gap}).
Although not obviously seen for stars that are more massive than K dwarfs in \cite{McQuillan2014}, \cite{Davenport2017} and \cite{Davenport2018} suggested that this gap might extend to G dwarfs by combining the Kepler period sample with Gaia data.
Using data from the MEarth project \citep{Berta2012},
\cite{Irwin2011} and \cite{Newton2016} did not see this intermediate period gap for stars with masses below the fully convective limit ($\sim$M3).
However, not enough rotation periods exist for stars at or beyond the fully convective limit to draw firm conclusions about the behavior of the gap at the late-M-dwarf end, until now.

Several explanations have been proposed to explain the intermediate period gap.
The most discussed hypotheses include a temporary pause in star formation in the local volume \citep[e.g.][]{McQuillan2013, Davenport2017}, detection bias caused by the low variability of light curves for stars in the gap \citep[e.g.][]{Reinhold2019}, and rapid spin-down following a temporary phase of stalled spin-down.
The stalled spin-down is thought to be caused by an onset of angular momentum transport from the radiative core to the convective envelope \citep{Spada2020}, which prevents the stellar surface from continuing its Skumanich-like spin-down \citep[see also][]{Curtis2020, Gordon2021}.
A period of rapid spin-down immediately after this (presumably also as a consequence of core--envelope interactions) would produce a physical gap. The star formation scenario has already been definitively ruled out by \citet{Curtis2020}: the intersection of cluster color--period sequences for 1.4~Gyr NGC~752 and 2.7~Gyr Ruprecht~147 intersecting the intermediate period gap at different colors means the gap cannot have been created by a single event in time. 

To disentangle the other two scenarios, period measurements for stars around the gap and close to the fully convective limit are needed. If the gap changes or disappears around the fully convective boundary, where stars no longer have a radiative core, the core--envelope interaction hypothesis will be favored.
However, periods for these stars are hard to measure not only because they are faint, but also because old M dwarfs can have periods $>$ 25 days. 
This means high precision, long-baseline photometric data are needed to expand the period sample.
One of the facilities that can provide such data is the Zwicky Transient Facility (ZTF).
ZTF is a ground-based optical time-domain survey that uses the Palomar 48~inch Schmidt telescope with a wide-field camera that has a 47 deg$^2$ field of view \citep{Bellm2019}.
Observing targets as faint as $\sim$ 19 mag for the entire northern sky, ZTF has produced billions of archival light curves with a typical cadence of once per day in three different optical filters \citep[$g, \sim$ 400--580\ nm; $i, \sim$ 580--700\ nm; $r,\sim$ 700--900\ nm;][]{Masci2019}.
With more than three years of observation data from ZTF, we are now able to measure long rotation periods for faint, low-mass stars. 
Period catalogs using ground-based telescopes have already shown success in reproducing the \kepler\ period distribution and measuring rotation periods for fully convective stars \citep[e.g.][]{Briegal2022,Newton2016,Irwin2011}.

With the goal of expanding available rotation period measurements towards lower masses in order to calibrate gyrochronology and understand the spin-down of low-mass stars, we present a new period catalog consisting of 40,553 late-K to M dwarfs using ZTF, in which 9,535 have periods $>$ 10 days. 
In section~\ref{sec:datamethod}, we describe the target selection, pipeline, and vetting criteria used to obtain the rotation periods.
In section~\ref{subsec:pd}, we compare our period distribution with those of \kepler\ and MEarth.
In section~\ref{subsec:gap} through~\ref{subsec:jump}, we discuss where the intermediate gap closes and the evaluate the different hypotheses for the gap in light of this new catalog.
In section~\ref{discussion}, we inspect how stars spin-down in light of the new period catalog.
We conclude in section~\ref{sec:conclusion}.

\section{Data/Method}\label{sec:datamethod}
\subsection{Target selection from \gaia}\label{subsec:data}
We selected stars from \gaia\ eDR3 \citep{gaia2020} with the following criteria:
\begin{itemize}
    \item Late K to M dwarfs: 
    \begin{itemize}
    \item \verb|bp_rp| $>$ 1 (temperature $\lesssim$ 5200~K)
    \item absolute magnitude in \gaia\ $G$-band $M_G>$ 7 (before accounting for reddening and extinction)
    \end{itemize}
    
    \item Observational constraints for ZTF \citep{Curtis2020}. Stars brighter than $G < 13$ are likely to saturate the ZTF CCDs, and stars fainter than $G>18$ are likely to be too noisy.
    \begin{itemize}
    \item 13 $<$ \verb|phot_g_mean_mag| $<$ 18
    \item \verb|dec| $> 0$
    \end{itemize}
    
    \item Quality flags taken from \cite{gaia2018HR}. 
    \begin{itemize}
    \item \verb|ruwe| $<$ 1.4
    \item \verb|parallax_over_error| $>$ 10
    \item \verb|phot_g_mean_flux_over_error| $>$ 50
    \item \verb|phot_rp_mean_flux_over_error| $>$ 20
    \item \verb|phot_bp_mean_flux_over_error| $>$ 20
    \item \verb|phot_bp_rp_excess_factor| $<$ \\ 1.3 + 0.06\;(\verb|bp_rp|)$^2$
    \item \verb|phot_bp_rp_excess_factor| $>$ \\ 1.0 + 0.015\;(\verb|bp_rp|)$^2$
    \item \verb|visibility_periods_used| $>$ 8
    \end{itemize}
\end{itemize}
    
\begin{figure}
    \centering
    \includegraphics[width=0.48\textwidth]{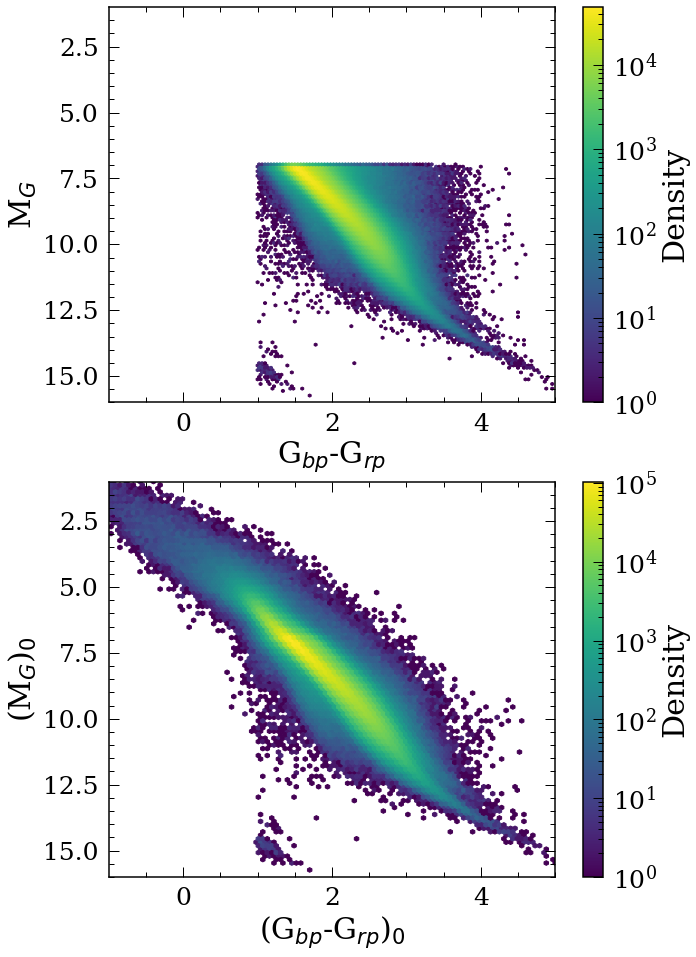}
    \caption{Color-magnitude diagram (CMD) for the $\sim$ 7 million selected stars of interest from \gaia\ eDR3.
    The top plot shows the raw \gaia\ absolute magnitude and color, and the bottom plot shows those after accounting for reddening and extinction with the Bayestar19 3D dust map \citep{Green2018, Green2019}.}
    \label{fig:1}
\end{figure}

We also excluded any data points with ZTF quality flags indicating poor conditions or data products, and selected only stars with more than 100 ZTF observations after applying the quality cuts.
This left us with $\sim$ 7 million stars.
We accounted for reddening and extinction from interstellar dust for each star using the \texttt{Bayestar19} 3D dust map \citep{Green2019} implemented in the \texttt{dustmaps} Python package \citep{Green2018}.
Figure~\ref{fig:1} shows the color--magnitude diagram (CMD) for our targets, colored by density.
The top panel shows the raw \gaia\ absolute magnitude and color, and the bottom plot shows those after accounting for reddening and extinction.

\subsection{Reduction pipeline}
We developed a simple pipeline to improve upon the archival light curves from ZTF and measured rotation periods for low-mass stars using the Lomb--Scargle (LS) periodogram.
To remove non-astrophysical, systematic signals that are common to multiple stars, our pipeline first retrieves surrounding archival light curves within a certain radius and magnitude of the targeted star.
It then removes the systematics by subtracting the median normalized flux of all the surrounding light curves from that of the targeted star at each timestamp. 
A similar method has been proven successful in \cite{Curtis2020}, in which they apply the same procedure to measure rotation periods for stars in the 2.7 Gyr old open cluster, Ruprecht 147, using the Palomar Transient Factory. 
Our method is applied directly to the archival light curves, rather than the image files, which saves a significant amount of computation time and storage.
We are able to improve on the photometric scatter of the original light curves by an average of 26\% using this pipeline, estimated using 1 million randomly selected light curves.
We measured this improvement by phase folding the original and the processed light curves onto their detected periods, splitting the folded light curve into 30 bins in time and calculating the average scatter across all bins.

After de-trending the archival light curves, we used the LS periodogram (\verb|astropy.timeseries.LombScargle|) from \verb|astropy| \citep{astropy:2013, astropy:2018} to measure the periods.
We measured periods for the entire light curve and for each season ($\sim$ 300 days per season) separately.
We set the maximum period detection to be 200 days for the periodogram since detecting periods over that is difficult with individual ZTF seasons being $\sim$ 300 days.
The ``best period'' for each star was defined to be the period with the highest normalized LS peak power (either from the entire light curve or from one of the seasons).
We then applied the vetting criteria to the best periods to obtain stars with correct period measurements (these vetting criteria are described in the next section).

\begin{figure*}
    \centering
    \includegraphics[width=0.9\textwidth]{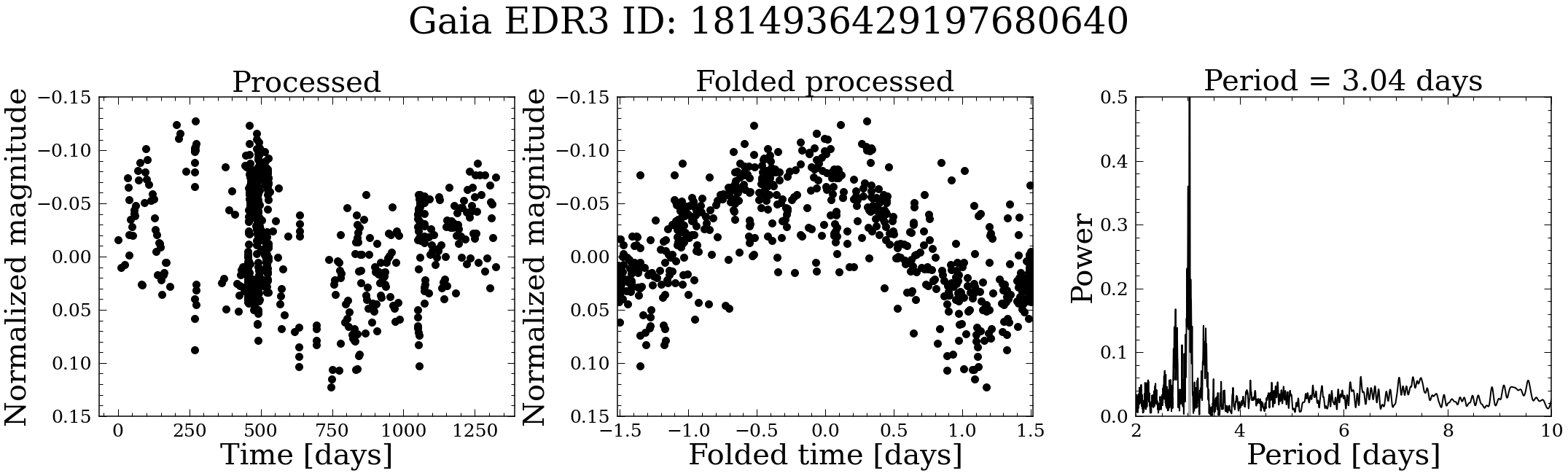}
    \includegraphics[width=0.9\textwidth]{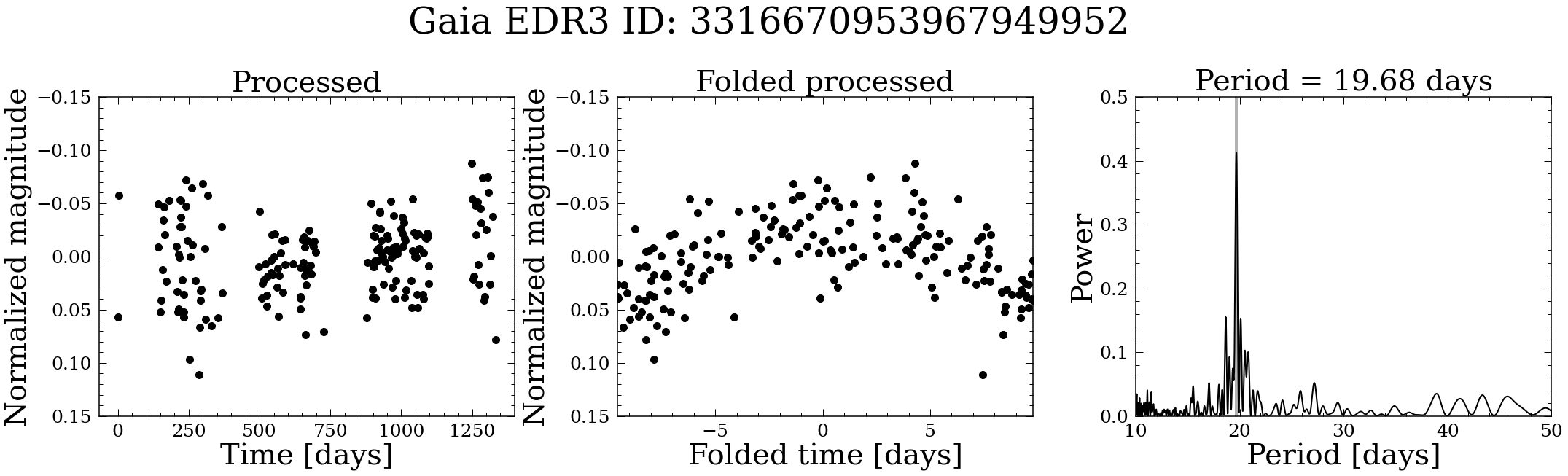}
    \includegraphics[width=0.9\textwidth]{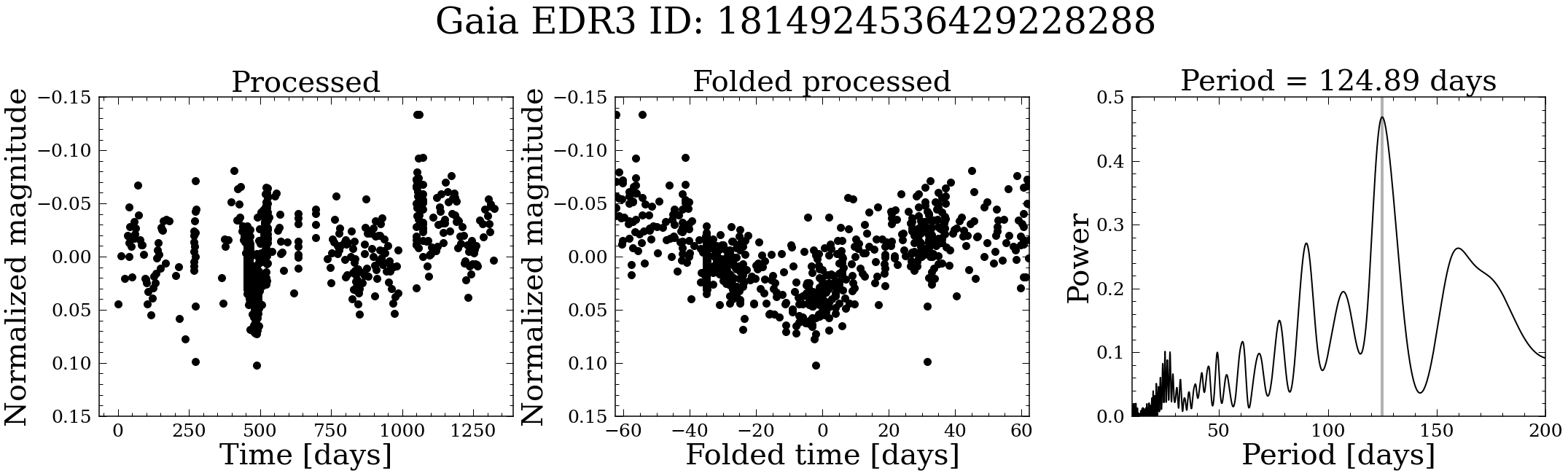}
    \caption{Three examples of the processed ZTF light curves (left column), processed ZTF light curves folded onto the detected periods (middle column), and the LS periodograms run on the entire light curve (right column) in different period ranges.
    The top row shows an example of \prot\ $<$ 10 days; the middle row shows that of 10 days $<$ \prot\ $<$ 100 days; and the bottom row shows that of \prot\ $>$ 100 days.}
    \label{fig:lc_example}
\end{figure*}

Figure~\ref{fig:lc_example} shows three examples in different period ranges of the processed ZTF light curves (left column), processed ZTF light curves folded onto the detected periods (middle column), and the LS periodograms run on the entire light curve (right column).
The top row shows an example of \prot\ $<$ 10 days; the middle row shows that of 10 days $<$ \prot\ $<$ 100 days; and the bottom row shows that of \prot\ $>$ 100 days.

\subsection{Optimizing the pipeline and vetting criteria to measure periods from MEarth}
We optimized the detrending algorithm to measure existing M dwarf rotation periods from the MEarth-North catalog \citep{Newton2016}.
This catalog contains 387 rotators (classified as grades ``A" and ``B" for convincing signals; grades ``U'' and ``N'' for uncertain and non-detections are not used here) with rotation periods ranging from 0.1 days to 150 days.
We tuned our pipeline to maximize the agreement between the measured ZTF rotation periods and the MEarth periods, as MEarth contains reliable period measurements for the slowly rotating low-mass stars we are most interested in.

First, we optimized the search radius and magnitude criteria used to select the neighboring stars required to calculate the systematics correction for our ZTF detrending pipeline.
We did this by maximizing the LS peak power for 
ten random stars in the MEarth sample. 
We performed a grid search over search radius between 1 and 5 arcmin in 1 arcmin increments, and relative $M_G$ between $\pm$0.5 mag to $\pm$4 mag of the target star in 0.5 mag increments.
The results suggest that the maximum LS peak power is not very sensitive to the search parameters, but selecting stars within 4 arcminutes and 2 magnitudes of the targeted star still provides the optimal result, which we therefore adopted for our pipeline. 

\begin{figure}
    \centering
    \includegraphics[width=0.48\textwidth]{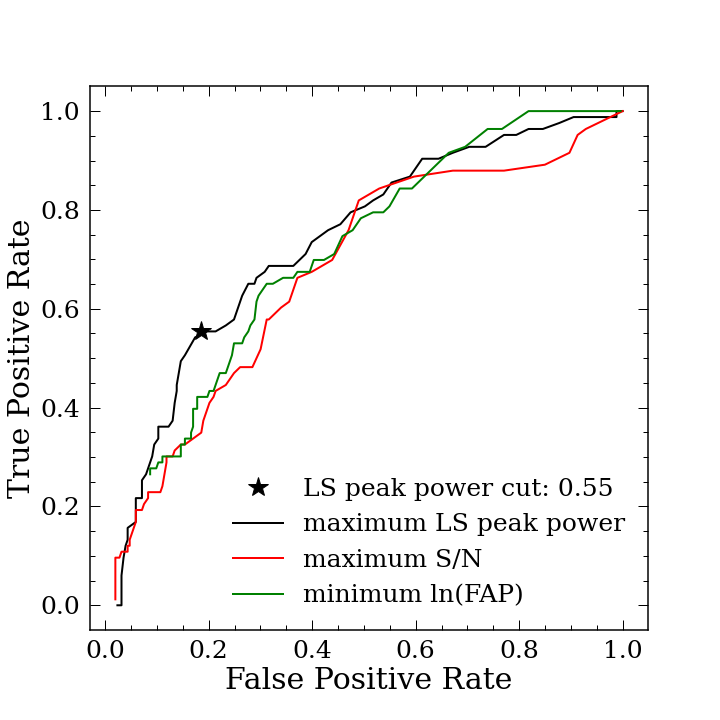}
    \caption{Receiver operating characteristic (ROC) curve for the three period indicators (black: maximum LS peak power; green: minimum ln(FAP); red: maximum S/N).
    Since our goal was to minimize the FPR while keeping the TPR high, we used the LS peak power to vet our periods.
    We accepted all periods that have LS peak powers $>$ 0.55 (black star), and this gives us a FPR of 19\% and a TPR of 55\%.
    We also used a $S/N$ cut of 10 to ensure the LS peak is significant. }
    \label{fig:2}
\end{figure}

We then optimized the period vetting criteria. 
To do this, we measured the 239 MEarth periods with $M_G$ between 13 and 17 using our ZTF pipeline and performed various cuts on three period indicators, namely the maximum LS peak power, maximum $S/N$ (calculated by the ratio of LS peak power to median LS peak power), and the minimum false alarm probability for the highest peak (FAP, provided by \verb|astropy|), to determine the best single vetting parameter that provided the lowest false detection rate while keeping most of the correct periods.
We wanted to first determine one vetting parameter that would exclude most incorrect periods and then fold in other parameter(s) to fine-tune the pipeline.
Figure~\ref{fig:2} shows the Receiver Operating Characteristic (ROC) curve for the three period indicators (black: maximum LS peak power; green: minimum ln(FAP); red: maximum $S/N$). 
The ROC curve is commonly used to test the performance of a classifier in machine learning.
The $x$-axis shows the False Positive Rate (FPR), calculated by, FPR = (\# of false positive)/[(\# of false positive) + (\# of true negative)];
and the $y$-axis shows the True Positive Rate (TPR), calculated by, TPR = (\# of true positive)/[(\# of true positive) + (\# of false negative)].
A ZTF period is identified as a good period (positive) when the measurement is within 10\% of the literature value.
The lines are created by plotting the FPR and TPR at different period indicator thresholds.
For example, we first identify positive periods with the criteria of maximum LS peak power $>$ threshold.
We can then calculate the FPR and TPR for that threshold.
By repeating this calculation with different threshold values, we can see how the FPR and TPR changes.
A perfect classifier would have a ROC curve that goes straight up from the bottom left corner (FPR = 0; TPR = 0) to the top left corner (FPR = 0; TPR = 1) and across to the top right corner (FPR = 1; TPR = 0).
This means, at a certain threshold (e.g. LS peak power $>$ some value), the perfect classifier can identify all periods that are correct and discard all incorrect periods. 
Since our goal was to minimize the FPR while keeping the TPR high, we selected the maximum LS peak power to vet our periods.
We accepted periods that have maximum LS peak powers $>$ 0.55 (black star), which gives us a FPR of 19\% and a TPR of 55\%.
In order to make sure the LS peak is significant, we also performed an $S/N$ cut of 10.\footnote{This $S/N$ cut is different than the maximum $S/N$ cut in Figure~\ref{fig:2}, as this $S/N$ is calculated for the (one season or the entire) light curve that contains the highest LS peak power. }

\begin{figure*}
    \centering
    \includegraphics[width=\textwidth]{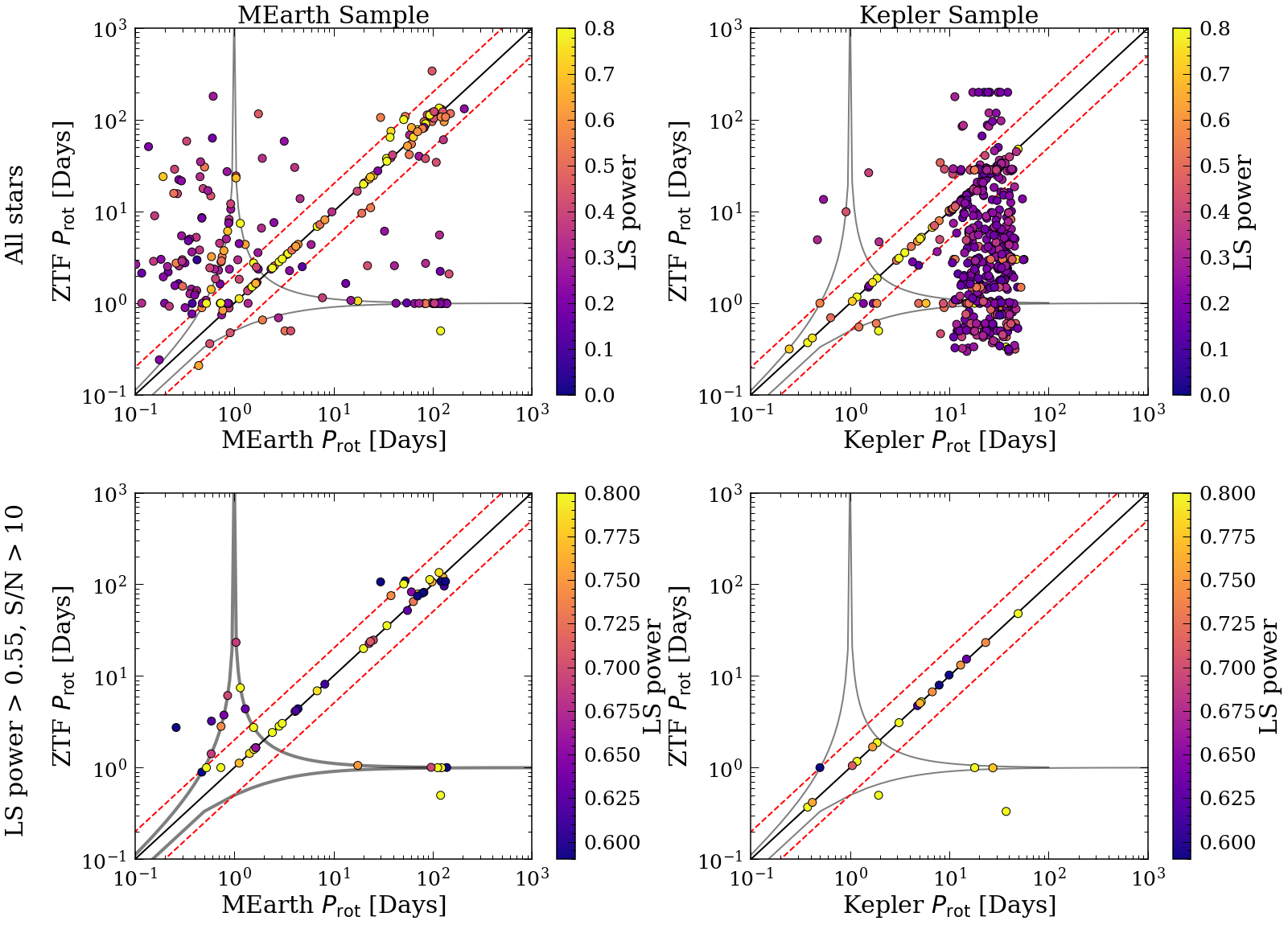}
    \caption{Performance of the vetting criteria (maximum LS peak power $>$ 0.55 and $S/N$ $>$ 10) for MEarth (left) and \kepler\ (right).
    The top row shows the period comparison for the un-vetted sample and the bottom row shows that of the vetted sample. 
    Stars are colored by their maximum LS peak powers.
    The black lines show the 1-to-1 line, and the red lines show the half-period harmonics. 
    The grey lines show the 1-day systematic alias calculated by $|1/P_{\rm rot} \pm 1|^{-1}$.
    The vetting technique is able to eliminate most disagreeing periods outside of the 1-day alias systematic. 
    The resulting TPR and FPR are 0.46 and 0.15 for MEarth, and 0.21 and 0.01 for \kepler.}
    \label{fig:3}
\end{figure*}

To test the reliability of our vetting technique, we also measured the rotation periods of 500 random \kepler\ stars using our ZTF pipeline. 
We then compared the measured periods to those published in \cite{McQuillan2014}.
Figure~\ref{fig:3} shows the vetting results.
Each column shows the un-vetted ZTF periods (top) and the vetted ZTF periods (bottom) versus the literature periods taken from MEarth \cite[left;][]{Newton2016} and Kepler \citep[right;][]{McQuillan2014}.
Stars are colored by their maximum LS peak powers.
The black lines show the 1-to-1 line, and the red lines show the half-period harmonics. 
The grey lines show the 1-day systematic alias calculated using $|1/P_{\rm rot} \pm 1|^{-1}$.
Typically, for even sampling time-series, the systematic will be exactly at the sampling cadence, which for ZTF would be at 1-day.
However, due to the uneven sampling, these `failure modes' will appear at $|m/P_{\rm rot} + n/\delta P|$, where $P_{\rm rot}$ is the true period, $\delta P$ is a period of max power in the window function's periodogram, which for this specific case is 1 day, and $m$ and $n$ are integers \citep[for more details, see,  e.g.][]{VanderPlas2018}.
It is clear that the vetting criteria, maximum LS peak power $>$ 0.55 and $S/N$ $>$ 10, are able to eliminate most disagreeing periods outside of the 1-day alias systematic. 
The resulting TPR and FPR are 0.46 and 0.15 for MEarth, and 0.21 and 0.01 for \kepler.

\subsection{Measure and vet ZTF periods for all targeted stars}
\begin{figure}
    \centering
    \includegraphics[width=0.48\textwidth]{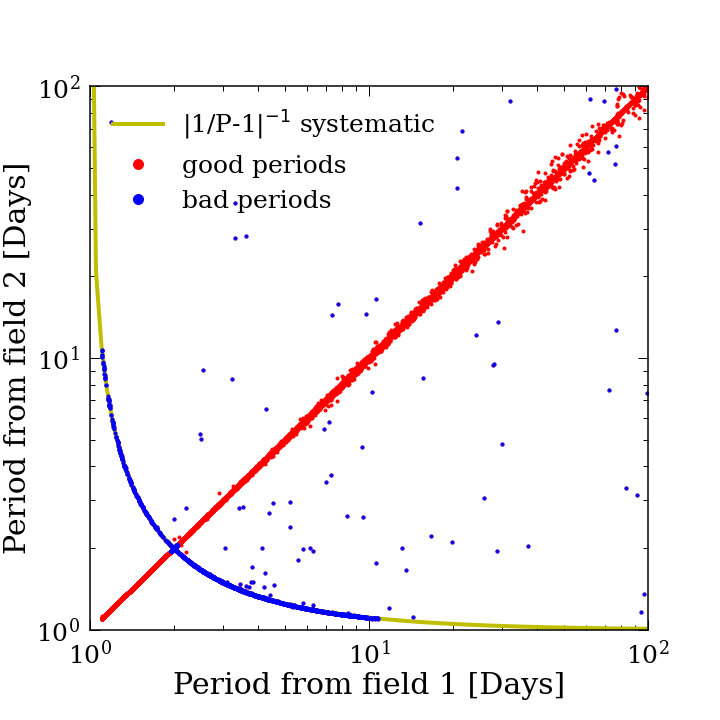}
    \caption{Period measured for the same stars but from separate ZTF fields/CCDs.
    All measurements have \prot\ $>$ 1.1 day, maximum LS peak power $>$ 0.55, and $S/N$ $>$ 10.
    The yellow line shows the 1-day systematic alias.
    Stars in blue are excluded from our catalog (either on the systematic line or do not agree within 20\%). 
    We take the period with the highest LS peak power to be the measured period for each star.
    }
    \label{fig:recover}
\end{figure}

After optimizing the pipeline and vetting criteria, we measured rotation periods from the ZTF archival light curves by first cross-matching the stars in ZTF with our \gaia\ target list (see section~\ref{subsec:data}), using an angular separation of 1.2 arcsec.
We found many potentially erroneous period measurements at \prot\ $\lesssim$ 1 day caused by the sampling alias \citep[also seen in][]{Briegal2022}. As a result, we only select stars with period measurements $>$ 1.1 days. 
Most stars in our \gaia\ list have two different period measurements from two separate ZTF fields/CCDs.
We can use these separate measurements to further characterize potential systematic signals in ZTF light curves. 
Figure~\ref{fig:recover} shows the result of plotting periods from two fields against each other for stars with \prot\ $>$ 1.1 day, maximum LS peak power $>$ 0.55, and $S/N$ $>$ 10 for both measurements.
Most periods measured from separate CCDs agree with each other. 
The systematic caused by the 1-day sampling alias is indicated by the yellow line.
To increase the reliability of our catalog, we only select stars that do not lie along the systematic line, and those with two periods agreeing within 20\%. 
We take the period with the highest normalized LS peak power to be the measured period for each star.
We want to warn the readers that ground-based systemics are more complicated due to the uneven sampling rate \citep[see also][]{Briegal2022, Irwin2011}.
We assumed periods that pass the vetting criteria and with agreeing measurements from different ZTF fields/CCDs are correct.

Even with this fairly aggressive vetting procedure, we are still left with 40,553 stars with good periods, in which 9,535 have measured periods $>$ 10 days.
This new period catalog greatly expands the number of measured periods for K and M dwarfs around and below the fully convective limit, allowing us to study stellar spin-down for low-mass stars, and to understand the nature of the intermediate period gap. 

\subsection{Comparison with previous measurements}
\begin{figure}
    \centering
    \includegraphics[width=0.48\textwidth]{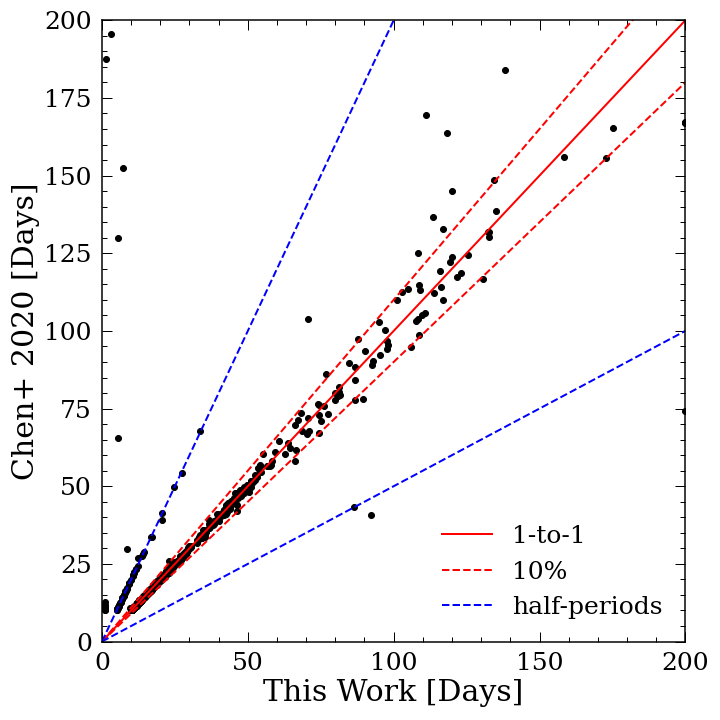}
    \caption{Period comparison for the same stars between \cite{chen2020} and this work.}
    \label{fig:chen22}
\end{figure}

\cite{chen2020} studied variable stars using ZTF DR2 data, in which they measured 150,000 periods (with $\sim$ 5,500 $>$ 10 days) using an LS periodogram applied to the entire light curve archive.
They vetted stars based on the false alarm probability ($<$ 0.001) and the goodness of fit ($R^2 <$ 0.4) of the fourth-order Fourier functions to the light curves.
We found 823 stars with period $>$ 10 days in common with period measurements from their study. 
The low number of overlapping targets is most likely caused by a) ZTF DR2 did not include the entire eventual ZTF footprint, b) the baseline of ZTF DR2 is only 400 days compared to 1,200 days for DR10, and c) our vetting cuts are more stringent.
Figure~\ref{fig:chen22} shows the periods measured from this work against those from \cite{chen2020}.

In comparison, most measurements agree within 10\%.
Some half-period aliases are also present in the sample, which is common and expected in rotation period studies as light curves often show large signals in their LS periodogram at half of the true periods.

\section{Results}\label{sec:result}
In this section, we  compare the ZTF period distribution with that from \kepler\ and MEarth in section~\ref{subsec:pd}.
We then discuss the behavior of the gap at the fully convective limit in section~\ref{subsec:gap}.
Hypotheses for the gap are then speculated upon in light of this new dataset in section~\ref{subsec:physics}.
The column descriptions for the resulting period catalog of 40,553 stars are provided in Table~\ref{tab:catalog}, and the data are available in a machine-readable format in the online journal.
The column descriptions for the full LS periodogram outputs for the entire sample are shown in Table~\ref{tab:catalog_all} and the data is available on Zenodo.
We also provide the processed light curves for stars with maximum LS power $>$ 0.55, and they can also be access through Zenodo (\url{https://zenodo.org/record/7131051#.YzebTC1h1KM}).

\begin{table}
\centering
\caption{Catalog description of the 40,553 vetted ZTF periods.
This table is published in its entirety in a machine-readable format in the online journal.}
\begin{tabular}{crcrcr}
\hline
\hline
Column & Unit & Description\\
\hline
\texttt{source\_id} & & \gaia\ eDR3 source ID\\
\texttt{Prot} & days & measured period \\
\texttt{ra}& deg & right ascension from \gaia\ eDR3\\
\texttt{dec}& deg & declination from \gaia\ eDR3\\
\texttt{bprp}& mag & \bprp\ from \gaia\ eDR3\\
\texttt{abs\_G}& mag & absolute magnitude from \gaia\ eDR3\\
\texttt{maxpower}& & normalized maximum LS power\\
\texttt{parallax}& 1/kpc & parallax from \gaia\ eDR3\\
\hline
\end{tabular}
\label{tab:catalog}
\end{table}

\begin{table}
\centering
\caption{Catalog description of the $\sim$ 7 million ZTF periods measured.
This table as well as the processed light curves for stars with maximum LS power $>$ 0.55 are published in its entirety on Zenodo (\url{https://zenodo.org/record/7131051\#.YzebTC1h1KM}).}
\begin{tabular}{p{0.2\linewidth}p{0.1\linewidth}p{0.55\linewidth}}
\hline
\hline
Column & Unit & Description\\
\hline
\texttt{source\_id} & & \gaia\ eDR3 source ID\\
\texttt{Prot\_all} & days & measured period from the entire light curve \\
\texttt{Prot\_1} & days & measured period from the first season light curve \\
\texttt{Prot\_2} & days & measured period from the second season light curve \\
\texttt{Prot\_3} & days & measured period from the third season light curve \\
\texttt{Prot\_4} & days & measured period from the fourth season light curve \\
\texttt{maxpower\_all}& & normalized maximum LS power from the entire light curve\\
\texttt{maxpower\_1}& & normalized maximum LS power from the first season light curve\\
\texttt{maxpower\_2}& & normalized maximum LS power from the second season light curve\\
\texttt{maxpower\_3}& & normalized maximum LS power from the third season light curve\\
\texttt{maxpower\_4}& & normalized maximum LS power from the fourth season light curve\\
\texttt{ra}& deg & right ascension from \gaia\ eDR3\\
\texttt{dec}& deg & declination from \gaia\ eDR3\\
\texttt{bprp}& mag & \bprp\ from \gaia\ eDR3\\
\texttt{abs\_G}& mag & absolute magnitude from \gaia\ eDR3\\
\texttt{parallax}& 1/kpc & parallax from \gaia\ eDR3\\
\texttt{ruwe}&  & ruwe number from \gaia\ eDR3\\
\hline
\end{tabular}
\label{tab:catalog_all}
\end{table}

\subsection{Period distribution}\label{subsec:pd}
\begin{figure}
    \centering
    \includegraphics[width=0.4\textwidth]{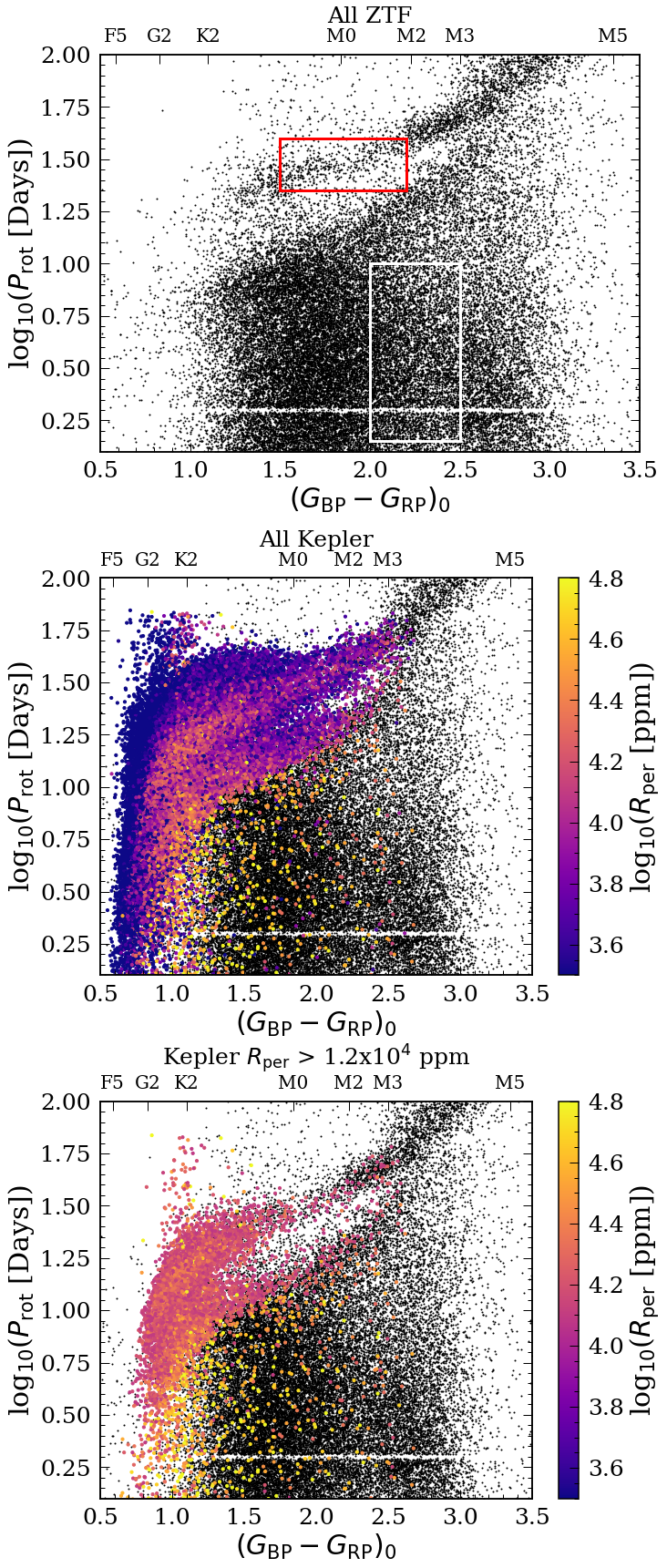}
    \caption{Period distribution of all vetted rotation periods from this work (top) compared to all periods from \kepler\ (middle) and to periods with light curve variability (\rper) $>$ 1.2$\times10^4$ ppm (bottom).
    The boxes show the corresponding features described in the text.
    The period distribution of ZTF matches well with \kepler\ stars with high variability. 
    The underdensity at \prot\ $\sim$ 2 days results from the systematic removal of stars on the 1-day alias line (yellow line in Figure~\ref{fig:recover}). 
    The underdensity of the distribution for stars right above the period gap (log$_{10}$(\prot) between 1.35 and 1.6 dex and \bprp\ between 1.5 and 2.2 dex; red box), if real, can provide insights on the formation theories of the gap.}
    \label{fig:kepler}
\end{figure}

Figure~\ref{fig:kepler} shows the vetted periods from this work (top) compared with \kepler\ periods taken from \cite{McQuillan2014} (middle/bottom).
The top panel shows the scatter plot for the vetted ZTF periods versus (\bprp)$_0$.
Stars with \prot\ $\sim$ 2 days were removed when we removed stars that lie on the 1-day alias line (see Figure~\ref{fig:recover}). 
The \kepler\ stars are colored by \rper, which is a measurement of the variability of the light curve.
\rper\ is computed by dividing the light curve into sections that span one rotation period, and by calculating the average 5\ith--95\ith\ percentile flux range across sections.
Rotation periods are not detectable when stars display little variability in their light curves, so \rper\ is a proxy for `detectability'.
Kepler is more sensitive than ZTF, so rotation periods can be detected from Kepler light curves for stars with lower \rper\ than ZTF light curves.
Comparing the ZTF distribution to the colored points of the full \kepler\ sample (middle plot), we notice the ZTF sample is missing many stars with low \rper.
Since ZTF is a ground-based telescope, which is heavily affected by the Earth's atmosphere, causing a higher noise floor, we expect to only recover periods for stars that have high variability.
Moreover, using a high maximum normalized LS power threshold to vet these periods also removes low-variability rotators.
After selecting \kepler\ stars with \rper\ $>$ 1.2$\times10^4$ ppm, we are able to reproduce the ZTF period distribution for the upper envelope (bottom panel).

We also observe a few new features, either real or caused by observational bias, that are not in the \kepler\ period distribution: 
\begin{enumerate}
    \item An underdensity of slowly rotating stars at $1.35 < \log_{10}(\prot) < 1.60$~dex and $1.5 < (\bprp)_0 < 2.2$~mag (red box in Figure~\ref{fig:recover} top plot).
    \item An overdensity of fast rotating stars with periods $<$ 10 days, especially for stars with $(\bprp)_0 < 2$~mag.
    \item An underdensity of fast rotating at $ 2.0 < (\bprp)_0 < 2.5$~mag (white box in Figure~\ref{fig:recover} top plot).
\end{enumerate}

The underdensity in 1), if real, lies right above the intermediate period gap, and could be caused by low variability for stars above the gap.
This could provide insights into the formation theories of the gap.  
The overdensity of fast rotation periods in 2) could be caused by an undetected systematic from the 1-day sampling alias (see yellow line from Figure~\ref{fig:recover}).
To further test the reliability of our measured periods, we separated the stars into two groups, one with ZTF periods $>$ 10 days and another with those $<$ 10 days.
We then randomly selected 1,000 stars from each group and obtained their TESS light curves using \verb|unpopular| \citep{Hattori2022}, an implementation of the Causal Pixel Model de-trending method to obtain TESS Full-Frame Image light curves.
The even sampling and high cadence rate of TESS facilitates the accurate measurement of short periods, and can be used to vet long ZTF periods that are affected by the 1-day alias.
The erroneous ``long'' ZTF periods that are affected by the 1-day systematic alias (these periods will appear to be at 1-day if the sampling rate is constant) can appear to be long due to the functional form of the ``failure modes'' created by the low cadence, 
$|1/P_{\rm rot} \pm 1|^{-1}$, as described in section~\ref{sec:datamethod}. 
As a result, if a star with ZTF period $>$ 10 days has a clear short period detected by TESS, the ZTF period is likely incorrect, and is created from the 1-day systematic alias\footnote{The moon's orbital period, $\sim$ 28 days, can also cause a 28-day systematic alias in ZTF data. However, since this does not seem to be significant in our sample, we will assume most periods are affected by the 1-day systematic alias.}.
To test stars with ZTF periods $<$ 10 days, we simply treat the TESS periods with high normalized LS power as ground truth and test how many periods ZTF can recover.

The results show for stars with measured ZTF periods $<$ 10 days, $\sim$ 50\% of their TESS periods with TESS normalized LS power $>$ 0.2 agree with the ZTF periods within 10\%.
Since periods measured with TESS can also be affected by systematics, we visually inspected 100 TESS light curves folded on the measured ZTF and TESS periods.
We chose to inspect TESS light curves as they should not be affected by the same ``failure modes'' and should be ideal for detecting rapid periodic signals.
From this analysis, we confirmed that $\sim$ 50\% of the stars with measured ZTF periods $<$10 days are most likely incorrect.
For stars with measured ZTF periods $>$ 10 days, we only discovered $\sim$ 5\% of the stars that have significant (TESS normalized LS power $>$ 0.2) short TESS periods ($<$ 10 days), meaning TESS only revealed $\sim$ 5\% of the measured long ZTF periods are potentially affected by the 1-day systematic.
To summarize, we estimate that 50\% of ZTF rotation periods measured to be $<$ 10 days are incorrect, and may often be longer than 10 days, and 5\% of ZTF periods measured to be $>$ 10 days are erroneously measured, short rotation periods, meaning the true periods for this sample are mostly likely shorter than measured from ZTF.
This test has shown the potential of combining TESS and ZTF. 
Future work can include a more sophisticated method to combine the photometry data from both surveys, and ultimately, loosen the strict vetting criteria for the ZTF periods applied in this paper.
Considering the high contamination rate for ZTF periods $<$ 10 days, we will only focus on stars with ZTF periods $>$ 10 days for the rest of the analysis.

We do not understand the cause of 3); however, as 50\% of these periods appear to be incorrect, this under-density may not be physical.

\begin{figure*}
    \centering
    \includegraphics[width=0.95\textwidth]{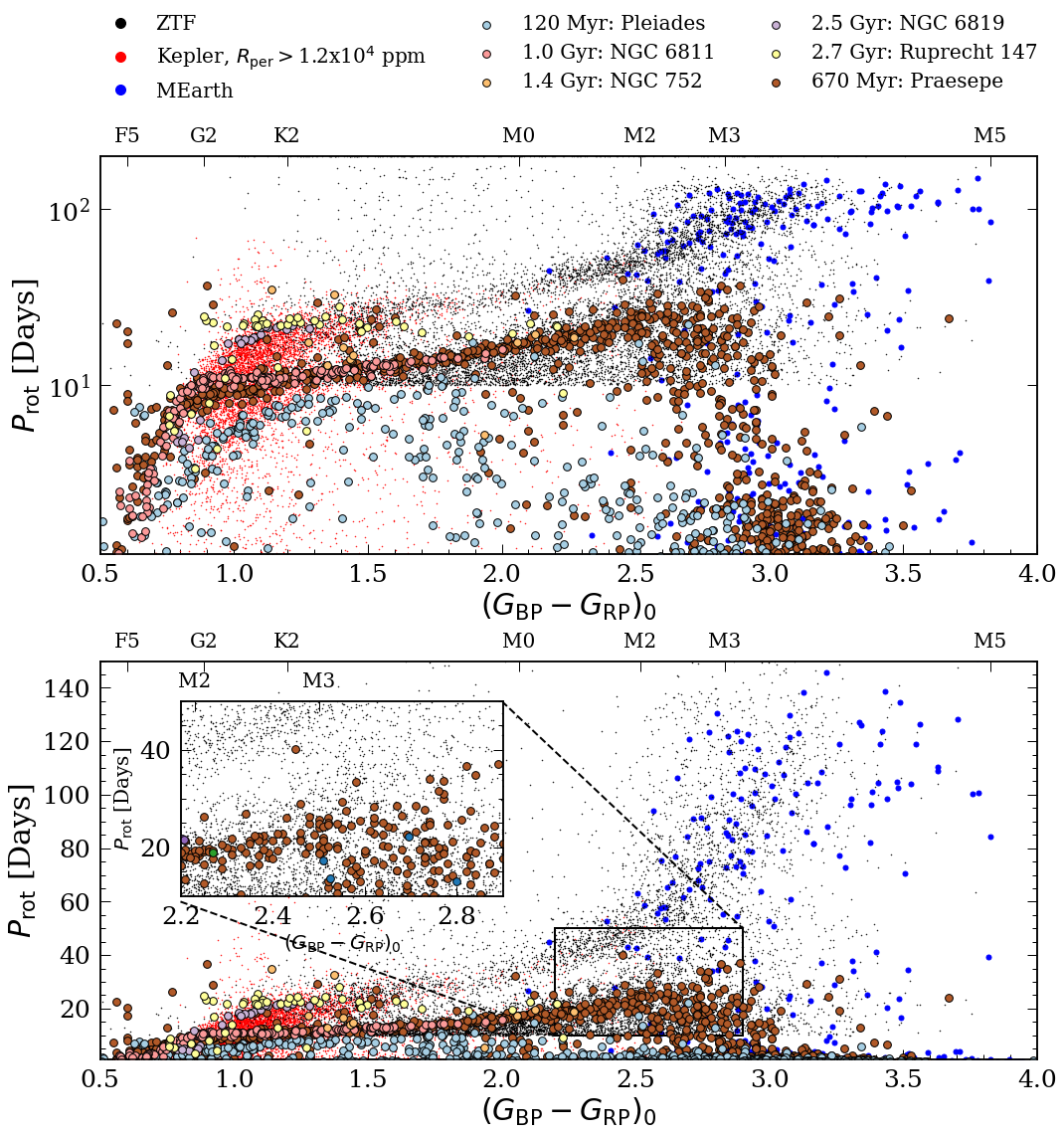}
    \caption{Color--\prot\ diagram for ZTF stars (black), MEarth stars (blue), \kepler\ stars with \rper\ $>$ 1.2$\times 10^4$ (red), Praesepe periods from \cite{Rampalli2021}, 
    and other cluster periods from \cite{Curtis2020} in log period space (top) and in linear period space (bottom). 
    The subplot in the bottom panel shows the de-reddened color versus rotation periods in linear space for ZTF and cluster stars around the fully convective limit.  
    The ZTF period distribution is able to bridge the gap between \kepler\ and MEarth for the first time.
    Based on the cluster data, we propose that stars beyond the fully convective boundary directly converge onto the slowly rotating sequence because they do not go through stalled spin-down as they do not have radiative cores.
    }
    \label{fig:totaldis}
\end{figure*}

With periods from ZTF, we are finally, for the first time, able to bridge the gap between \kepler\ and MEarth.
Figure~\ref{fig:totaldis} shows the dereddened color--\prot\ diagram in log period space (top) and in linear period space (bottom) for ZTF ($>$ 10 days; black), MEarth (blue), and \kepler\ stars with \rper\ $>$ 1.2$\times 10^4$ (red). Also shown are data from benchmark star clusters including Praesepe \citep[700~Myr][]{Rampalli2021}, 
and other clusters tabulated in \citet{Curtis2020}.\footnote{\citet{Curtis2020} gathered \prot\ data for the Pleiades \citep[120 Myr;][]{Rebull2016}, Praesepe \citep[not shown here;][]{Douglas2019}, NGC~6811 \citep[1 Gyr;][]{Curtis2019}, NGC~752 \citep[1.4 Gyr;][]{Agueros2018}, NGC~6819 \citep[2.5 Gyr;][]{Meibom2015}, and Ruprecht~147 \citep[2.7 Gyr;][]{Curtis2020}, then crossmatched the sample with Gaia~DR2, corrected colors for interstellar reddening, and filtered out candidate binaries.} The subplot in the bottom plot shows the de-reddened color versus rotation periods in linear space for ZTF and cluster stars around the fully convective limit. 

Looking at the overall period distribution, the ZTF period distribution matches nicely where it overlaps with those from \kepler\ and MEarth.
A few key observations include: 
\begin{enumerate}
    \item The gap in color--period space may close at the fully convective limit.
    This limit occurs around 0.35$\pm$0.05~\msun\ \citep[e.g.][see section~\ref{subsec:gap} for more details]{Feiden2021,Chabrier1997,Dotter2008,vanSaders2012,Chen2014}.
    \item The period distribution for low-mass stars beyond the fully convective limit flares up and shows a gap between the fastest and slowest rotating stars.
    \item There is a fanning out of the slowly rotating sequence for stars beyond the fully convective limit.
\end{enumerate}

More specifically, 2) describes the rapid increase in rotation periods for slowly rotating M dwarfs beyond the fully convective limit, and 3) describes the increase in the range of detected periods for the slowly rotating M dwarfs with increasing color (see the bottom panel of  Figure~\ref{fig:totaldis}; \bprp\ between 2.5 to 3.5, periods between 60 to 150 days for both features).
Both 2) and 3) are also pointed out in the MEarth sample by \cite{Newton2016} and \cite{Irwin2011}.
A study of the kinematics of MEarth stars suggests the large gap between the slow and fast rotating late M dwarfs (distinct from the `intermediate period gap' studied here) is caused by a rapid spin-down of these stars \citep{Newton2016}.

We compare cluster data with our new field star rotational distribution in figure \ref{fig:totaldis}. Similarly to what was found in \cite{Curtis2020}, stars in clusters younger than 2 Gyr that are converged onto the slowly rotating sequence appear to lie along the lower edge of the gap.
The 2.7 Gyr Ruprecht~147 cluster contains stars that span the gap: its G and early-K dwarfs lie above the gap and its late-K and early-M dwarfs lie below it.
Because the K and M dwarfs in Ruprecht~147 have rotation periods similar to the K and M dwarfs in the younger, 0.67 Gyr Praesepe cluster, \citet{Curtis2020} theorized that these low-mass stars are potentially experiencing a temporary stalling in their spin-evolution  \citep[presumably due to core--envelope coupling; see also][]{Spada2020}.
The fact that stars in Ruprecht~147 are clearly undergoing stalling {\it paired} with the fact that Ruprecht~147 crosses the gap suggests that the mechanism responsible for this apparent stalling might be related to the formation of the gap.

Core--envelope coupling is the leading hypothesis for the stalled spin-down seen in the clusters, particularly Ruprecht~147, a cluster in `mid stall'.
If the gap is also related to core--envelope coupling (which seems likely given that Ruprecht~147 spans it), we can test whether  core-envelope coupling is the cause of stalling by examining the rotation periods of fully convective stars.
If we assume that fully convective stars are unlikely to experience the same kind of core--envelope decoupling and recoupling that stars with distinct cores and envelopes might, they should directly converge onto the slowly rotating sequence without stalling.
This scenario seems to be supported by these new data: the intermediate period gap appears to close at the fully convective boundary.

\subsection{Where does the gap disappear?}\label{subsec:gap}
\begin{figure*}
    \centering
    \hspace*{-1cm}\includegraphics[width=0.83\textwidth]{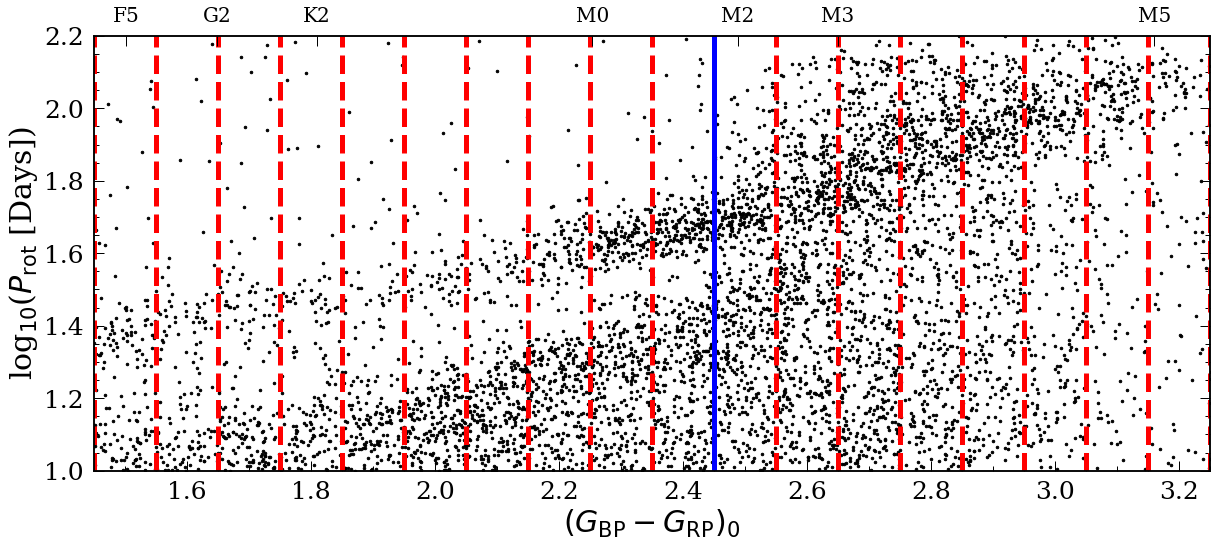}
    \hspace*{-0.7cm}\includegraphics[width=0.835\textwidth]{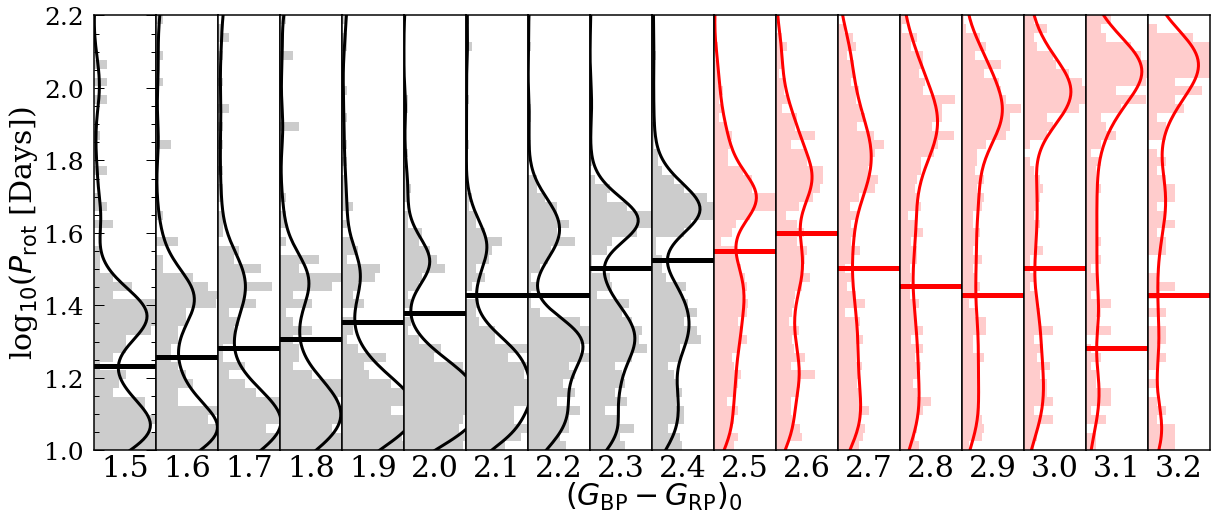}
    \includegraphics[width=0.88\textwidth]{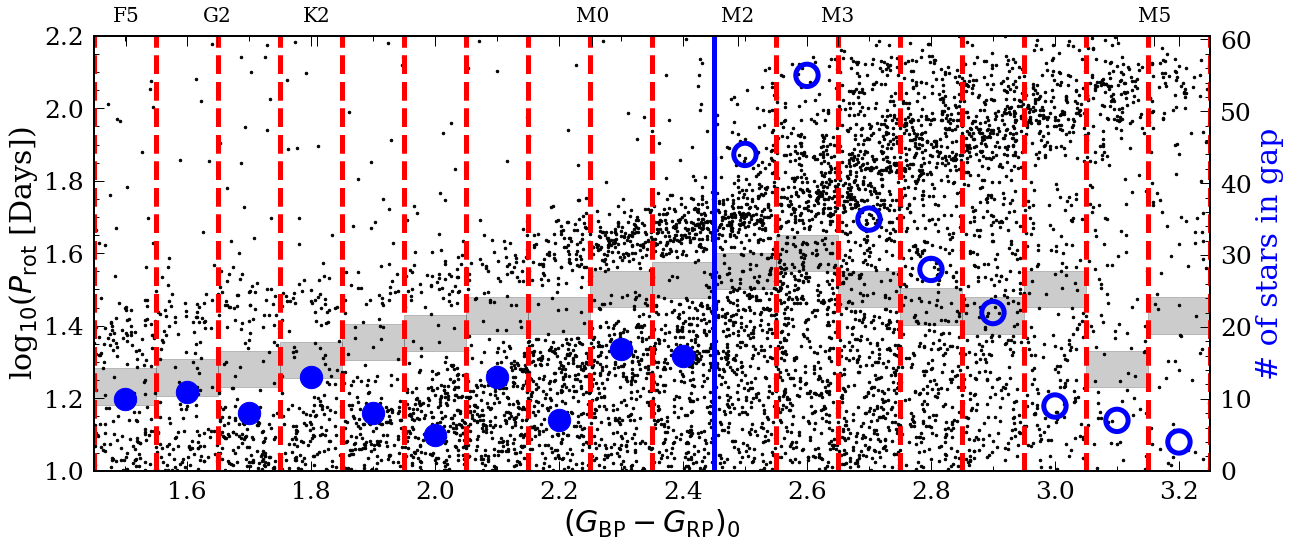}
    \caption{{\it: Top}: The period distribution for stars with \prot\ $>$ 10 days and low reddening (Av $<$ 0.2).
    The red dashed lines show the edges of the bins in the bottom plot.
    The blue solid line (\bprp\ = 2.45) shows the color of the left bin edge for the first bin without the gap based on the number of stars in the gap for each color bin.
    {\it Middle}: Period histograms for stars in 0.1 mag \bprp\ bins (from 1.45 mag to 3.25 mag).
    The solid lines show the KDE for each sub-population.
    The horizontal lines show the local minimum determined using the KDEs.
    These lines indicate the locations of the intermediate period gap in each color bin (when there is one).
    The colors of the histograms and lines indicate the bins with the intermediate period gap (black), and those without (red).
    {\it Bottom}: Same as the top plot but plotting the number of stars (blue points) in the gap for each color bin.
    Empty circles indicate the bins without a noticeable gap.
    The numbers are determined by calculating how many stars lay within 0.05 dex of log10(\prot) around the local minimuma (grey boxes).
    The trend suggests the number of stars in the gap stays relatively constant from \bprp\ = 1.5 mag to 1.7 mag and the gap eventually closes up around \bprp\ of 2.45 mag.}
    \label{fig:gap}
\end{figure*}

To determine the color where the intermediate period gap closes up, we examine in Figure~\ref{fig:gap} the stars around the gap with \prot\ $>$ 10 days and low interstellar reddening to sharpen up the feature (we selected stars with low visual extinction obtained from Bayestar19: $A_V < 0.2$~mag).
The top panel shows \pgbr$_0$ versus $\log_{10}(\prot)$ for this subsample.
The middle plot shows the histograms of periods for stars in narrow color bins that are 0.1 mag wide, ranging from 1.45 mag to 3.25 mag \citep[this method is also used by][]{McQuillan2014}.
The solid lines overlaying each histogram are the Kernel Density Estimates (KDEs), calculated using \verb|seaborn.kdeplot| with a smoothing parameter of 0.2.
Horizonal solid lines show local minima determined by finding the minimums between the two highest peaks in the KDE curves.
These lines indicate the location of the intermediate period gap  (when present) in each color bin.
The red dashed lines show the edges of the bins in the bottom panel.
The bottom panel shows the same data as the top panel, and adds the number of stars in the gap for each color bin (blue points);
empty circles indicate the bins lacking a noticeable gap.
The numbers are determined by calculating how many stars lay within 0.05 dex of $\log_{10}(\prot)$ around the local minima (grey boxes).
The trend suggests that the number of stars in the gap is relatively constant between $1.5 < \pgbr_0 < 1.7$~mag and the gap eventually closes up around $\pgbr_0 \approx$ 2.45~mag (blue solid line),
where the number of stars in the gap rises sharply.
The number of stars in the gap (grey boxes) decreases again at larger colors/lower masses, where the gap location is incorrectly identified using the local minimum of the KDE (because there is no significant gap).
The colors of the histograms and the lines in the middle panel indicate the bins with a clear gap (black; where the absolute number of stars in the gap is low) or and those without (red; where the absolute number of stars in the gap is high, or where the gap location is not accurately determined).

To determine whether the gap closes at the fully convective boundary, we estimated the color at which the boundary starts using MESA Isochrones \& Stellar Tracks \citep[MIST;][]{Choi2016}.\footnote{Available at \url{https://waps.cfa.harvard.edu/MIST/interp_isos.html}}
To get the color for the fully convective boundary, we need to first understand how metallicity can affect this value, as we do not have accurate abundance information for the M dwarf sample.
Since stars with \prot\ $>$ 10 days and $A_V < 0.2$~mag (the sample shown in Figure~\ref{fig:gap}) are located very close to the Sun (within 500 pc based on \gaia\ photometry), the average metallicity of our sample is likely to be close to $-$0.2 dex \citep[][assuming most stars are low-$\alpha$ stars and have Galactic heights $<$ 1 kpc]{Schlesinger2014}.
The MIST isochrone of age 1~Gyr indicate the theoretical temperature for the fully convective boundary occurs around 3,400--3,600 K if we vary the metallicity between $-$0.3 dex and 0.0 dex.
The color range for this temperature range can be estimated with the relation described in \cite{Curtis2020} to be around 2.2--2.5.
The \bprp\ value of where the gap disappears lays within this color range.
As a result, we suggest that the gap likely disappears at or near the fully convective boundary.

Even though the gap vanishes close to the fully convective limit, there is still an overdensity of stars at the slowly rotating end, indicating convergence of fully convective stars onto the slowly rotating sequence.
This phenomenon is also pointed out in \citep{Newton2016, Irwin2011,Popinchalk2021}

\subsection{Can uncertainties in period measurements make the gap appear to close?}
Sun-like stars rotate at different rates at latitudes ranging from their equators to their poles. Assuming that at any time, active regions are concentrated at a particular latitude, measuring drifts in the periods derived from long-term photometric monitoring can be attributed to this latitudinal differential rotation \citep[e.g.,][]{Henry1995, Donahue1996}. 
In the case of our ZTF sample, we currently have data for 2--3 seasons, which is not yet sufficient for definitively detecting differential rotation. However, we do measure seasonal differences, which might be caused by a combination of differential rotation and measurement uncertainties that increase with period duration. 

\begin{figure}
    \centering
    \includegraphics[width=0.48\textwidth]{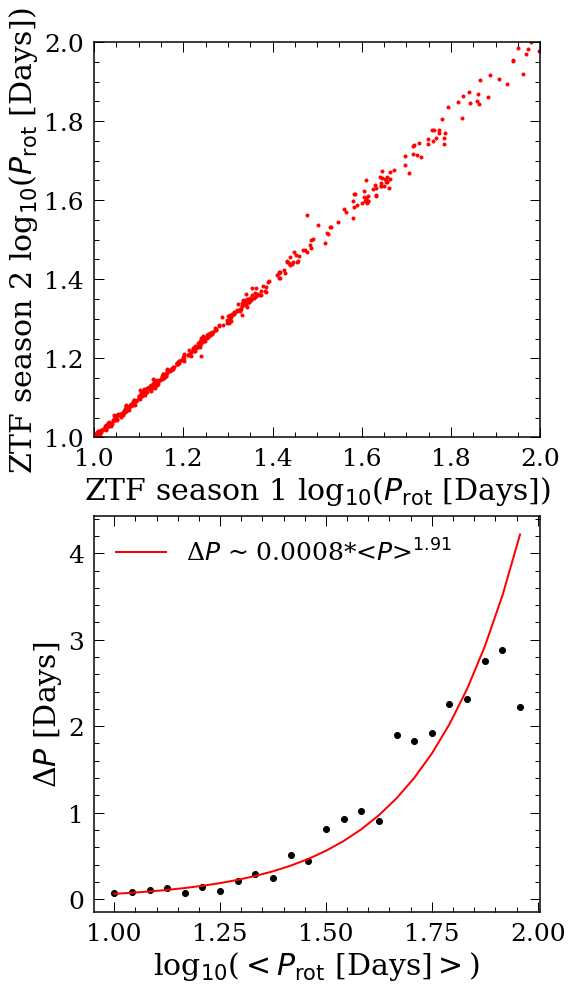}
    \caption{Measurement of period uncertainty, $\Delta P$. 
    This uncertainty is likely a combination of differential rotation and measurement uncertainty.
    We calculate $\Delta P$ by first selecting all stars with at least two good period measurements (LS peak power $>$ 0.55 and $S/N$ $>$ 10) from two different seasons. 
    Stars with season 1 period and season 2 period between a small bin of $\log_{10}(\prot)$ are then selected to calculate the log average period, $\log_{10}(<$\prot$>$), and $\Delta P$.
    We do this by taking the mean and standard deviation of these periods in linear space, respectively. 
    We then estimated the the relation by fitting an exponential relation between $\Delta P$ and $\log_{10}(<$\prot$>$).}
    \label{fig:dp}
\end{figure}

While we cannot yet distinguish between the two, it is crucial to know whether the gap disappears at the slowly rotating limit simply because our measurement errors have increased.
In order to investigate this possibility, we calculate the period uncertainty by first selecting all the stars with at least two good period measurements (LS peak power $>$ 0.55 and $S/N$ $>$ 10) from separate seasons (see top panel of Figure~\ref{fig:dp}). 
We first bin the stars by their log$_{10}$(\prot) and color.
The bins are between 1 to 2 dex and are 0.04 dex wide.
In each bin, we select stars with both season 1 log$_{10}$(\prot) and season 2 log$_{10}$(\prot) within that bin.
The log of the average period, $\log_{\rm 10}(<\prot>)$, and the period uncertainty, $\Delta \prot$, is then calculated by taking the mean and standard deviation of the periods (in linear space) for stars in each bin.
The lower panel of Figure~\ref{fig:dp} shows that $\Delta \prot$ increases with \prot. 
We then describe this relation with  an exponential fit, finding $0.0008<P>^{1.91}$.
Using this relation, a 30-day period measurement would have a seasonal difference of $\sim$ 0.5 days in rotation, which is not large enough to close the gap. 
By perturbing the periods around the gap with low reddening (Av $<$ 0.2) within their assumed uncertainty, we find the gap persists (by examining at the KDE in small color bins; see also Figure~\ref{fig:gap}) until the uncertainty is $\sim$ 100 times more than estimated above or the error is 30\% of the measured periods, which is unlikely.

\subsection{Why does the gap close?}\label{subsec:physics}
The three leading hypotheses advanced to explain the gap include
1) a temporary pause in star formation in the local volume \citep[e.g.][]{McQuillan2013, Davenport2017}, 2) a detection bias caused by stars showing reduced variability in their light curves when they are in the gap \citep[e.g.][]{Reinhold2019}, and 3) a temporary epoch of stalled spin-down in low-mass dwarfs, possibly followed by a brief rapid spin-down across the gap \citep[e.g.][]{Curtis2020, Gordon2021}.
We can now reexamine each of these hypotheses with our new period catalog that covers periods for stars around the fully convective limit. 

For 1), this scenario was invoked to explain the gap when it was first discovered with \kepler.
Since \kepler\ only looked at a relatively small patch of the sky, and the gap laid between two gyrochronology isochrones in color--period space, a local pause in star formation seemed to be a reasonable explanation. 
However, many subsequent studies have ruled against this scenario.
First, \cite{Curtis2020} showed that the color--period sequence for the 2.7-Gyr-old Ruprecht~147 cluster does not run parallel to, but in fact intersects, the gap. This means the gap cannot be formed at a single age, as was previously thought.\footnote{As stalling had not been recognized at the time the gyrochronology isochrones used by \citet{Davenport2017} were created \citep{Meibom2011_M34}, they did not accurately describe how stars actually spin down.}
Furthermore, the persistence of the gap in all-sky surveys \citep[e.g.,][this work]{Gordon2021, Briegal2022} suggests this can only happen if the pause in star formation occurred on a large scale, throughout the local Galaxy, which has not been observed in other (non-rotation) age-dated samples. 
All recent studies rule strongly against this hypothesis.

Scenario 2), proposed by \cite{Reinhold2019}, suggests stars in the gap have a temporarily reduced variability due to the transition from spot-dominated to faculae-dominated photospheres, and thus, their rotation periods are difficult to detect using broadband photometric time series. 
This is an appealing argument as stars do have lower variability in and around the gap for the \kepler\ sample (see Figure~\ref{fig:kepler}). 
However, the cause of the decrease in variability is still unclear and our sample neither supports nor disagrees with this theory. 

For 3), \cite{Curtis2020} and \cite{Gordon2021} hypothesized that after a brief period of stalled surface spin-down due to core--envelope coupling (where the surface angular momentum lost to magnetic winds is countered by the momentum transfer from the faster-rotating core), low-mass stars resume spinning down and transition rapidly across the period gap.\footnote{Although the rapid spin-down is not physically understood.}
Our period distribution supports this explanation, as the period gap disappears right at the fully convective limit where the radiative core vanishes, and thus no core--envelope interaction can happen.

The gap does seem to be related to the stalling phenomenon: stars appear to be stalling when they are right beneath the gap in period-color space.
Chronologically speaking, a star's surface rotation stops decreasing (stalls), and whatever causes the gap happens immediately after that.
Either the star's variability decreases for a period of time, or the star's surface rotation slows dramatically and the star jumps across the gap.
Though possible, it would be a coincidence if the phenomena responsible for stalling and the gap happened at right around the same time/evolutionary stage/rotation periods but weren't related.
So, if a spot-dominated to faculae-dominated transition, or other period of reduced variability, is responsible for the gap, it is logical to assume that this process is caused by stalling (core-envelope coupling).

Another possible scenario that could lead to a gap caused by a detection bias is the following: reduced internal differential rotation caused by core-envelope coupling could reduce the magnitudes of stars' magnetic fields and render them less photometrically variable for a period of time.
This would cause a gap created by a detection bias because stars with little photometric variability would not appear in rotation period catalogs.
This hypothesis could be supported by our new period distribution because the gap disappears at the fully convective limit, where core-envelope coupling cannot take place.
However, this scenario raises the question: why does stellar variability increase again?
Would that mean that core and envelope decouple once more?
These questions may be answered in future by investigating the evolution of stellar magnetism using new observations and simulations.

\subsection{Do stars evolve quickly across the gap?}\label{subsec:jump}
In this section, we investigate whether the gap is 1) physical (i.e., truly empty) or 2) caused by photometric detection bias.
For example, if the gap is physical, this can indicate stars go through a phase of rapid spin-down after the radiative core and the convective envelope are coupled.
On the other hand, if the gap is solely caused by detection bias, stars should continue spinning down continuously across the gap ``Skumanich style'' after their cores and envelopes are recoupled, and simply show low variability either because they become less magnetically active, or perhaps because they are transitioning from being spot- to faculae-dominated as suggested by \citet{Reinhold2019}.

As pointed out in \cite{Santos2021}, they are able to measure stars with lower variability and expand the upper envelope of the period sample in \cite{McQuillan2014}. 
However, they do not find more stars in the intermediate period gap compared to \cite{McQuillan2014}, suggesting the gap could be truly empty.
If the gap is caused by detection bias, the ability to measure periods for stars with lower variability should result in a narrower gap, which is not what is found in \cite{Santos2021}.

Another way to distinguish these two scenarios is by looking at the kinematic ages for stars below and above the gap. 
If stars indeed jump over the gap, kinematic ages for stars above the gap will be very similar to those just below, as they would have spun down on a very short timescale.
However, this method only works if enough precise, unbiased kinematic ages are available for stars around the gap, as stars will also only take $\sim$ 400 Myr to evolve across this gap if they are spinning down $\propto t^{1/2}$.

We performed a preliminary test with \kepler\ gyro-kinematic ages taken from \cite{Lu2020}. 
These ages are kinematic ages estimated using an age-velocity dispersion relation taken from \cite{Yu2018}.
The velocity dispersion for each star is calculated from the vertical velocities for stars that are similar in periods, temperatures, and absolute magnitude to the targeted star.

\begin{figure}
    \centering
    \includegraphics[width=0.48\textwidth]{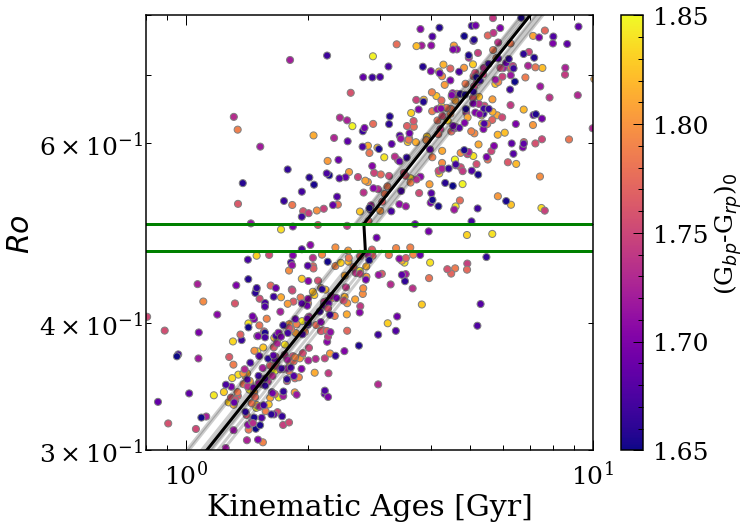}
    \caption{Rossby number, \ro, versus gyro-kinematic ages for \kepler\ for stars with \bprp\ between 1.65 mag and 1.85 mag in \cite{Lu2020} with \prot\ $>$ 10 days, colored by \bprp.
    We picked this color range since the gap is most prominent. 
    The green lines show the gap in constant \ro\ of 0.47 dex and 0.50 dex.
    The gray lines show the linear fits, assuming Skumanich spin-down, in every 0.02 mag color bin.
    We are only fitting the intercepts, $b$, by using log(\ro) = 0.5\,log(age)+$b$.
    The fits are done separately for each color bin below and above the gap.
    The black lines show the average relation from the individual fits.
    The discontinued jump between the black lines below and above the gap suggests stars could have gone through a phase of rapid spin-down after core--envelope re-coupling.
    However, more rigorous testing is needed to confirm this idea.
    }
    \label{fig:gyroage}
\end{figure}

Figure~\ref{fig:gyroage} plots the Rossby number, \ro, versus gyro-kinematic ages 
from \cite{Lu2020} for \kepler\ 
stars with \prot\ $>$ 10 days, 
and $1.65 < \bprp\ < 1.85$~ mag where the gap is most prominent. 
Rossby number is strongly correlated with chromospheric activity \citep{Noyes1984, Mamajek2008} and is defined to be \prot/$\tau_{\rm cz}$, in which $\tau_{\rm cz}$ is the convective turnover timescale. 
We estimated $\tau_{\rm cz}$ using eq.~36 from \cite{Cranmer2011}. 
The temperature is taken from \cite{Lu2020} and is derived with the \citet{Curtis2020} color--temperature relation using \bprp,  accounting for reddening with \verb|dustmap| \citep{Green2018}.
We use \ro\ instead of \prot\ in this test as the period gap seems to be relatively constant in \ro\ (between 0.47--0.50, indicated by the green horizontal lines). 
The gray lines show the linear fits, assuming Skumanich spin-down, in every 0.02 mag color bin; i.e., we are only fitting the intercepts, $b$, by using log(\ro) = 0.5\,log(age)+$b$.
The fits are done separately for each color bin below and above the gap.
The variance in the intercept values is around 3\%.
The black lines show the average relation from the individual fits.
The discontinued jump between the black lines below and above the gap suggests the stars could have gone through a phase of rapid spin-down and jumped over the gap after core--envelope coupling.

However, we do not claim from this test that stars ``jump over'' the gap.
More rigorous tests with more gyro-kinematic ages are needed to confirm this result.
By combining radial velocity measurements from \gaia\ DR3 in the near future, and current or future period catalogs, more kinematic ages can be obtained to study the gap and calibrate spin-down models. 

\section{Stellar spin-down based on observations}\label{discussion}
With our new ZTF period measurements, we can start to investigate stellar spin-down across the fully convective boundary.
A new ground-based survey of stellar rotation in M67 (4~Gyr) has produced rotation periods for K7--M3 members \citep{Dungee2022_talk}. Their preliminary results reveal a converged sequence of slow rotators extending just beyond the fully convective boundary, running just above the upper edge of the intermediate period gap, and along the lower edge of the population of slow rotators surveyed by MEarth \citep{Newton2016}. 
In detail, looking at Praesepe in Figure~\ref{fig:totaldis}, together with the new M67 results (not shown), and assuming stars in M67 do indeed lay on a common sequence, stars that are fully convective should spin down continuously from Praesepe (670 Myr) to M67 (4 Gyr) as the \prot\ distribution of stars are relatively smooth beyond the fully convective limit.
On the other hand, stars with radiative cores first go through a phase of stalled spin-down due to core--envelope coupling, and first pile up at the lower boundary of the gap.
After their cores and envelopes couple, they most likely evolve quickly over the gap and meet with stars that are fully convective and converge onto the same sequence indicated by M67.

\section{Conclusions}\label{sec:conclusion}
In this paper, we selected K and M dwarfs from \gaia, and measured their rotation periods using archival ZTF light curves that we refined using neighboring stars to define a systematics correction.
We vetted the periods by only selecting measurements with maximum LS peak power $>$ 0.55 and $S/N$ $>$ 10.
$S/N$ is estimated by dividing the maximum LS peak by the median LS peak power, and 
we calculated this $S/N$ for the season (or the entire light curve) that had the maximum LS peak power.
We also only selected stars with measurements agreeing within 20\% from different ZTF fields or CCDs that also did not lie on the 1-day sampling alias systematic. 
Our final catalog provides 40,553 period measurements (9,553 periods $>$ 10 days) that bridge the gap between the previous \kepler\ and MEarth period sample, allowing us to uncover the behavior of stars and the intermediate period gap around the fully convective boundary. 

From this sample, we noticed the disappearance of the intermediate period gap near the fully convective limit (\bprp\ $\sim$ 2.45). 
We tested whether seasonal differences in period measurements caused by differential rotation or measurement error can close the gap, and concluded it is unlikely. 
We discussed the three possible explanations for the period gap in light of this new catalog.
We were able to further rule out the explanation of a temporary pause in star formation and were in favor of a phenomenon connected to core--envelope coupling since the gap disappears right around the fully convective boundary.
By using gyro-kinematic ages, we also suggested stars rapidly spin-down across the gap. 
Overall, we favor a phase of rapid spin-down as the cause of the gap, however, more rigorous statistical tests with an expanded set of gyro-kinematic ages are needed to confirm this idea.

\begin{acknowledgments}

This work is funded by NSF grant AST-2108251.
J.L.C. is supported by NSF grant AST-2009840. 


This work has made use of data from the European Space Agency (ESA)
mission Gaia,\footnote{\url{https://www.cosmos.esa.int/gaia}} processed by
the Gaia Data Processing and Analysis Consortium (DPAC).\footnote{\url{https://www.cosmos.esa.int/web/gaia/dpac/consortium}} Funding
for the DPAC has been provided by national institutions, in particular
the institutions participating in the Gaia Multilateral Agreement.
This research also made use of public auxiliary data provided by ESA/Gaia/DPAC/CU5 and prepared by Carine Babusiaux.


This research was done using services provided by the OSG Consortium \citep{OSG1,OSG2}, which is supported by the National Science Foundation awards \#2030508 and \#1836650.

This research has also made use of NASA's Astrophysics Data System, 
and the VizieR \citep{vizier} and SIMBAD \citep{simbad} databases, 
operated at CDS, Strasbourg, France.

\end{acknowledgments}

%

\vspace{5mm}
\facilities{Gaia, PO:1.2m \citep[ZTF;][]{ztfdata,ztftime}}


\software{  \texttt{Astropy} \citep{astropy:2013, astropy:2018},
            \texttt{Astroquery} \citep{astroquery},
             \texttt{dustmaps} \citep{Green2018},
            Matplotlib \citep{matplotlib}, 
            NumPy \citep{Numpy}, 
            Pandas \citep{pandas}, 
            \texttt{TESScut} \citep{Astrocut},
            \texttt{tess\_cpm} \citep[beta version from 2019; the published version is now called \texttt{unpopular};][]{Hattori2022},            
            }






\bibliography{sample631}{}
\bibliographystyle{aasjournal}



\end{document}